\documentclass[aip,reprint,superscriptaddress,eqsecnum]{revtex4-1}
\usepackage{hyperref}
\usepackage{bm,bbm}
\usepackage{amsfonts}
\usepackage{graphics,graphicx}
\usepackage{framed}
\usepackage{amsmath,amssymb}
\usepackage{dcolumn}
\usepackage[mathscr]{eucal}
\usepackage[toc,page]{appendix}
\usepackage{lipsum}
\usepackage{mathrsfs}
\usepackage{appendix}
\usepackage[usenames,dvipsnames]{xcolor}
\usepackage{cancel}

\begin{document}

\preprint{APS/123-QED}


\title{Influence of weak reversible cross-linkers on entangled polymer melt dynamics}

\author{Mohau J. Mateyisi}
\email[]{mateyisi@aims.ac.za}
\affiliation{Leibniz Institute of Polymer Research, D-01069 Dresden, Germany}


\author{Jens-Uwe Sommer}
\affiliation{Leibniz Institute of Polymer Research, D-01069 Dresden, Germany}

\author{Kristian K. M\"{u}ller-Nedebock}
\thanks{}
\affiliation{Institute of Theoretical Physics, Department of Physics, Stellenbosch University, Private Bag X1, Matieland 7602, South Africa}

\author{Gert Heinrich}
\thanks{}
\affiliation{Leibniz Institute of Polymer Research, D-01069 Dresden, Germany}

\date{\today}

\begin{abstract}
In this paper we study a system of entangled chains that bear reversible cross-links in a melt state. The cross-links are tethered uniformly on the backbone of each chain. A slip-link type model for the system is presented and solved for the relaxation modulus. The effects of entanglements and reversible cross-linkers are modelled as discrete form of constraints that influence the motion of the primitive path. In contrast to a non-associating entangled system the model calculations demonstrate that the elastic modulus has a much higher first plateau and a delayed terminal relaxation. These effects are attributed to the evolution of the entangled chains as influenced by tethered reversible linkers. The model is solved for the case when linker survival time $\tau_s$ is greater than the entanglement time $\tau_e$ but less than the Rouse time $\tau_R$.

\end{abstract}
\maketitle 

Reversibly associating polymers have been extensively studied over the past two decades from both
theoretical and experimental point of view (see, \emph{e.g.},~refs.~\onlinecite{leibler1991,tanaka1992viscoelastic} and references therein). This study is motivated by a recent experimental development on physical polymer networks in which a commercially
available bromobutyl rubber (BIIR) which is converted to a reversibly cross-linking material without permanent cross-links \cite{Das2015}. This is achieved by transforming bromine functional groups in BIIR into imidazolin
bromine groups. When the material is damaged, temperature induced rearrangement of ionic groups
enables the material to reorganise the ionic bonds leading to recovery of the mechanical and dynamical
properties. 

What makes this type of polymer system interesting, from applications point of view, is the possibility of tunability of its dynamic properties, such as stress relaxation, that is brought about by the existence of tethered reversible ionic
groups whose properties, such as connecting tether length, stiffness and linker bond strength, can be varied. This type of physical network has a potential to lead to the
development of a new class of smart materials that are more durable with an ability to fully recover
their mechanical properties at ambient temperature following damage.

A large body of literature has been dedicated to understanding the contributions of the associating components to the physical properties of entangled systems \cite{Cates1988,RubinsteinANDSemenov2001,
RubinsteinANDSemenov1998,Semenov2006,
vanRuymbeke2015}. It turns out that despite successes made in predicting dynamical and mechanical properties of such networks, 
there is a broad parameter space, that is determined by the nature of the associating components, which is largely unexplored. Associating groups as reported in the experiments by Das, \emph{et al.} \cite{Das2015} are responsive to temperature changes and they are tethered by side chains whose stiffness and length can be varied. The presence of tethers, to which the associating groups are encored to the parent chains, leads to a certain degree of movement for any pair of cross-linked segments. This is a new feature whose consequences merits an analytic treatment and this shall be one of the main focus of this paper. 
 
In this paper, we argue that upon deformation tethers can sustain stress and  that the association of linkers leads to a slowing down of the motion of the primitive chain segments in response to a deformation. We assume that tethers are of a length that allows cross-linked units to move within a distance of the order of the entanglement mesh size. 

Theories in which the life time $\tau_s$ of an associating sticky group is assumed to be long compared to all relaxation times, that it controls all relevant relaxation scales, are classified as kinetically limited theories \cite{Fredrickson2009}. Here we consider weak linker associations leading to short-lived reversible bonds, but also not necessarily saturated bonds in the context of a kinetically limited theory.  We demonstrate that this effect can be treated as an effective additional drag which depends on the parameters of the tethers and the reversible cross-links. In this way, we propose simple corrections to the Rouse time with explicit dependence on the physical parameters of the system, such as tether spring constant $\kappa$, number of chain segment $N$ and the respective kinetic rates $\omega_a$ and
$\omega_0$ of linker association and dissociation.

The mathematical treatment presented in this paper demonstrates that the evolution of the fraction of reversible cross-links that can sustain stress, upon deformation, can be modelled as some form of uniformly distributed  constraints.

Much progress has been made in predicting both viscoelastic and dynamical properties of dense entangled chain systems, in general, through the mean field tube model construction of de Gennes \cite{deGennes1971}, Doi and Edwards \citep{doiandEdwards1988} and, equivalently, through the slip-link variant of the tube model~\cite{doi2013}. The slip-link approach entails a simple picture in which some discrete ring-like polymer constraints are assumed to embody the other chains that restrict the wriggling of a reference chain to a small length scale, of the order of the mesh size $a$. In this tube model variant, short time scale motion of the polymer is represented by the  wriggling of the primitive chain. Motion on long time scales involves the primitive chain sliding along the slip-links leading to the creation and destruction of the slip-links upon visitation by chain ends. At intermediate positions along the chain, the number of slip-links is assumed to be fixed. The model reasonably captures the essence of the molecular motion in the dense system.

The primary objective here is to argue how the effects of weakly associating tethered reversible cross-links can be incorporated in the slip-link model with the aim of predicting the consequences of such effects on the relaxation modulus. We also discuss on how such a qualitative  model prediction can be used to interpret  experiments on linear viscoelasticity in weakly associating reversible networks. We derive that for reversible cross-links of short longevity, there is an effective friction on chain monomers leading to a slowing of the chain reptation dynamics. The relaxation of reptating chain ends released from entanglement constraints is also modified on short time scales by the necessity to relax short-lived cross-linkers there too. We conclude by comparing storage modulus curves.

It is worth mentioning that for quantitative 
comparison with experiments, one has to take into account many-body effects emanating from the motion of other chains relative to any reference entangled chain in the system.    
Other variants of the tube model for entanglements \citep{Read2008,graham2003microscopic,likhtman2002quantitative}, which take many-body effects into account, could potentially be modified to incorporate reversible cross-linking. An in-depth discussion on such effects as mediated by reversible cross-linking is interesting, in its own right, but it is outside the scope of this paper and will not be presented here.

The paper is organised as follows: In Section~\ref{Model description}  a modification to the tube model to incorporate reversible cross-linkers in the context of primitive path analysis is proposed. In Section~\ref{Linker Association dissociation induced monomeric friction} ideas on how the effective contribution of weakly associating cross-linkers can be analytically modelled as friction are presented.  Sections~\ref{primitive chain motion}, \ref{sec:release} and \ref{Stress Relaxation} deal with the relaxation of the chain at different times scales.
The paper is concluded with a discussion on new qualitative features of the elastic modulus.    
 
\section{Model description}
\label{Model description}

The reversibly cross-linked entangled system is made primarily of a melt of entangled linear polymers of length $L$. For simplicity each chain is assumed to be tethered by short side chains that are equally spaced. In addition to entanglements, reversible cross-links contribute additional constraints whose contribution on the dynamics depends on the energetics of the sticky groups, the length and stiffness of the connecting tethers. 

\begin{figure}[ht]
\centering
  \includegraphics[width=.46\textwidth, angle=0]{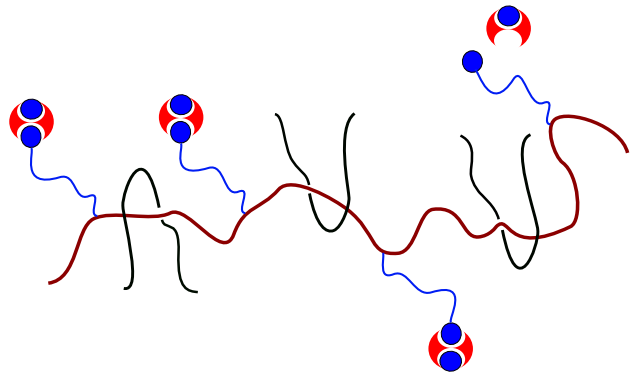}
	\caption{A depiction of a reference chain (brown) constrained by entanglements (shown in black) and reversible cross-links (blue). Reversible cross-links are tethered to the chain backbone through permanent cross-links. The ionic groups at the end of the tether may associate forming pairs of cross-links or even clusters of cross-links. For simplicity we will only model a system consisting of pairs of cross-links. }
	\label{tubemodel}
\end{figure}

On experimental studies the length of the side-chains can, in principle, be varied to any desired length. For the purpose of this discussion, reversible linkers are assumed to be tethered, as shown in FIG\ref{tubemodel}, to the parent chains by side chains of a length that is of the order of the entanglement mesh size. In this way the cross-linkers find their cross-linking partners within a distance of the order of the slip-link ring diameter or entanglement mesh-size diameter.

\begin{figure}[ht]
  \includegraphics[width=.5\textwidth, angle=0]{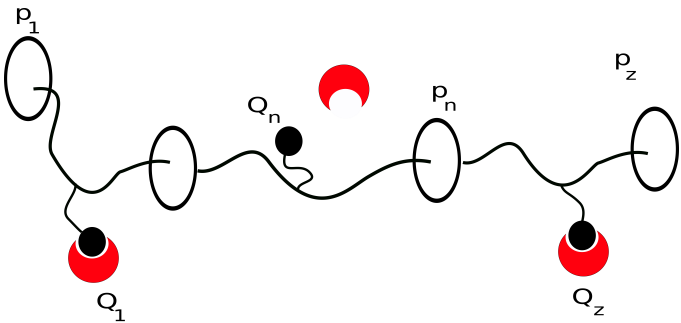}
	\caption{A slip-link model representation of a reversible entangled chain in a melt. Regions demarcated in red depict a possible cross-linking points while entanglements are represented by rings.}
	\label{slip-link model}
\end{figure}

As highlighted earlier, a slip-link model envisages an ideal situation in which an entangled chain is represented by an effective chain running through ring-like polymers or a polymer with loops. In a situation where chains bear reversible linkers, the model can simply be modified to incorporate the reversible linkers through some discrete tethered sticky points in between slip-links (See FIG.\ref{slip-link model}).
Reversible cross-links undergo association and dissociation reactions at the rate $\omega_a$ and $\omega_d$, respectively.
The dissociation rate depends on the extension of the tethers and temperature. As the chain end moves out of the slip-links, the assumption is that a slip-link is destroyed which is equivalent to a vanishing of the corresponding entanglement. The time evolution of the chain entails creation and release of slip-links at the chain ends and dissociation and association of reversible linkers. In this way the equilibrium number of cross-links and slip-links remains constant.

The discrete nature of the association and dissociation of linkers reactions makes the problem much more involved for stickers of long survival time $\tau_{s}$. As stated earlier the model presented here concerns linker survival times that are greater than entanglement time $\tau_{e}$ but much less than the Rouse time $\tau_{R}$. In this time scale the effect of weak linker association is simply a temporary deprivation of the kinetic energy to a diffusing monomer. The physical origin of the slowing down effect is the fact that the cross-linked sub-units have to drag along the connecting tethers as they explores space.
We anticipate that the case where reversible cross-link survival time is shorter than the entanglement time leads to a functional form of friction
which differs from characteristic molecular friction by simple pre-factors and this case will not be discussed here.
 
Let us consider a situation where the entangled chain system is deformed in time $t$ such that some sub-units of the chain are compressed while others are stretched. If we think of the center of mass of a primitive chain for a given test chain that is relaxing to recover an equilibrium length, we can say that it feels a velocity  dependent 
force that resists its motion due to the reversible linkers attachment.

On average the velocity dependent force that resists the motion of the primitive chain  $\bm{\Gamma}(\bm{v})$ can be expressed as a sum of two contributions
\begin{equation}
\bm{\Gamma}(\bm{\dot{r}})=
\bm{\Gamma}_f(\bm{v})+\bm{\Gamma}_{L}(\bm{v}),
\end{equation}
where $\bm{\Gamma}_f(\bm{v})$ is the usual force necessary to overcome friction from molecules brushing across each other and the solvent and is given by $\zeta_f \bm{v}$.
The second term captures the effect of reversible cross-linker attachment and it can be approximated to
\begin{eqnarray}
\Gamma_L(\bm{v}) =\bm{\Gamma}_L(\bm{v}=0)+ \bm{v} \left.\left(\frac{\partial \bm{\Gamma}_L}{\partial \bm{r}}\right)\right|_{\bm{v}=0} + \mathcal{O}(\bm{v}^2),
\label{tension expansion}
\end{eqnarray}
with the first term on the right equal to zero. This also approximates the effective tension on the reversible cross-linker tethers due to the chain motion, atleast within a cross-link survival time. The remaining terms of the expansion evaluate to $\bm{\Gamma}(\bm{v}=0)=0\text{ and }\bm{v} \left(\frac{\partial \bm{\Gamma}}{\partial \bm{\dot{r}}}\right)|_{\bm{v}=0}= \bm{v}\zeta_L$, where $\zeta_L$ denotes the reversible linker association-dissociation reaction induced friction coefficient (more details on this shall follow in 
section \ref{Linker Association dissociation induced monomeric friction}). 

The force needed to pull the reference primitive chain along the slip-links and reversible linkers per unit velocity is therefore $\zeta_c= \zeta_f + \zeta_L$. From the Einstein relation the diffusion of the center of mass can be expressed as
\begin{equation}
\label{center of mass diffusion coefficient}
D_c = \frac{k_{\text{B}}T}{ \zeta_f + \zeta_L},
\end{equation}
where $\zeta_f\sim N $ and $\zeta_L\sim N_x $.  This appears like a trivial modification of the diffusion coefficient of a Rouse model.
In a situation where there is
control on the adhesion strength of the reversible cross-linker \cite{Seiffert2013}, it is desirable to have a diffusion 
coefficient with an explicit
dependence on the kinetic rates. It is therefore necessary to express the friction coefficient correction term in terms of the kinetic coefficients of the corresponding attachment and detachment reactions with an appropriate molecular weight dependence. 

In order to approximate the total number of reversible cross-links that are in an associated state at equilibrium, we have to bear in mind that cross-linking happen when there is a certain degree of chain sub-units overlap in space. In addition, a cross-links must be energetically favourable. For the system at equilibrium, the number of cross-links $N_{\text{x}}$, that are in an associated state is given by
\begin{equation}
\label{number of associated linkers}
N_{\text{x}}\approx  I P_{\text{x-link}}.
\end{equation}
where $I= Z^2f^2$ is  the equilibrium total number of overlapping reversible linker bearing chain subunits, where $Z$ is the total number of chain sub-units bound by entanglement constraints and $f$ denotes the fraction of reversible linkers  in the system.
If we denote the molecular weight
of the reference chain $M$ and that of the chain in between entanglements $M_e$, then the number entanglements \cite{doiandEdwards1988}
\begin{equation}
Z= \frac{M}{M_e}.
\end{equation}

 The cross-linking probability $P_{\text{x-link}}$ is related to temporal probability $P_a$ of realising a cross-link, at a given time $t$ and spatial likelihood $P_s$ of  a pair of free linkers being within exploration volume $\delta^3$ of each other.
\begin{equation}
\label{cross linking probability}
P_{\text{x-link}}= \left\langle P_sP_a(t,\tau,s_0)\right\rangle=  P_s P_a,
\end{equation}
where an average is taken over all initial associated cross-linker positions $s_0$. The term $P_a(t,\tau;s_0)$ denotes the average probability distribution that a linker at chain position $s_0$ is found attached at time $t$ given that it has been attached for a duration $\tau$. It defines the temporal probability of reversible linker attachment.

From an argument developed by De Gennes for a related system \cite{deGennesI1982}, the spatial likelihood of a cross-link $P_{s}$ can be approximated by imagining the coils of two chains, whose volume is $R^3_0$ each, that intersect within a volume $V$ in space. The spatial probability that one sub-unit overlaps with a sub-unit of the other second chain is of the order of
\begin{equation}
P_s=\frac{V}{R^3}\frac{\delta^3}{R^3_0}.
\end{equation}
Since a chain is spread over a volume $R^3_0$, for the case where the coils have strong overlap $V\approx R^3_0$, this leads to
\begin{equation}
\label{association probability}
P_s\approx\frac{\delta^3}{R^3_0}.
\end{equation}
The temporal probability $P_a$ is much harder to calculate but for the system close to equilibrium it can be reasonably approximated by its steady state value. This is done in section \ref{Linker Association dissociation induced monomeric friction}.

\section{Linker association-dissociation kinetics}
\label{Linker Association dissociation induced monomeric friction}

As mentioned earlier, $P_a(t,\tau;s_0)$ denote the probability that at time $t$ a linker has been attached to the segment $s_0$ of the chain already for a duration $\tau$. We follow a similar scheme dealing with molecular motor attachment, shown by Banerjee, \emph{et al.}\cite{banerjee2011motor}.
We make the simplifying assumption that this function is independent of the position meaning that

\begin{equation}
P_a(t,\tau;s_0)=\frac{1}{N_l}P_a(t,\tau)
\end{equation}
here $N_l$ denotes the total number of reversible linkers tethered on the a reference entangled chain. 
The time evolution of the  probability $P_a(t,\tau)$ is given by
\begin{eqnarray}
\nonumber
\label{time evolution of attachment probability}
\lefteqn{
\partial_t P_a(t,\tau)+\partial_{\tau}P_a(t,\tau)}
\nonumber \\ & = & \omega_{a} \delta(\tau)P_{d}(t)-\left\langle \omega_d(\bm{\Delta} (\tau))\right\rangle P_a(t,\tau).
\end{eqnarray}
The term $P_d(t)$ is the probability that a cross-linker is detached at time $t$, while $\left\langle\omega_d(\bm{\Delta} (\tau)\right\rangle)$ and $\omega_a$ are the respective average dissociation and association rates. The attachment rate for a linker that has been attached for some time $\tau$ depends, in principle, on the extension of the tether $\bm{\Delta} (t,\tau)$. The probability distribution is normalized thus
\begin{equation}
\label{eq:PaNormalisation}
\int_{0}^{\infty}d\tau \int_{-N/2}^{N/2} d s_{0} P_{a}(t,\tau;s_0) + P_{d} =1.
\end{equation}

For simplicity, the probability $P_{a}(t,\tau;s_0)$ of linker attachment is assumed to be uniform and the linker tether is assumed to relax instantaneously to the relaxed state upon detachment. The detachment rate $\omega_d$ should depend on the degree to which the side chain is stretched. In equation (\ref{eq:PaNormalisation})
the detachment rate is 
replaced by its mean where an average for the rate is taken over all initial attachment positions of the linkers.

For related problems involving association and dissociation of tethered reactants, several forms of $\omega_d$  have been proposed in the literature \citep{banerjee2011motor}, \emph{e.g.} such as the exponential  $\omega_d=\omega_0 \exp(\alpha |\bm{\Delta}|)$ or the simpler form
\begin{equation}
\label{detachment rate}
\omega_d= \omega_0(1+ \alpha^2|\bm{\Delta}^2|).
\end{equation} 
The expressions for the rates are general and can also be used for the case of a non-active dissociation reactions. The terms $\alpha^{-1}$ and $\omega_0$ are the characteristic detachment length scale  and linker escape frequency, respectively. 
A common feature of these rates is that there is a characteristic extension beyond which a linker simply detaches. For simplicity, we shall adopt the quadratic form for the detachment rate. In this form of detachment rate the temperature and tether extension dependence is captured, to a lowest order, by the inverse length scale $\alpha^{-1}$ and $\Delta$ respectively.

\subsection{Mean field approximation}

As mentioned earlier, the survival time of a reversible cross-link depends on the amount of tether extension. Likewise, the tether extension depends on the linker survival time thus one can self-consistently define the mean field cross-link association waiting time  $t\approx\tau_{MF}$ as
\begin{equation}
\label{attachment time approximation}
\tau_{MF} = \left\lbrace\left\langle \omega_d \left[\bm{\Delta}(\tau_{MF})\right]\right\rangle
\right\rbrace^{-1}.
\end{equation}
The initial attachment position $s_0$ is taken to be independent of time. This allows us to introduce the mean field assumption 
\begin{equation}
\label{attachment probability approximation}
P_a(t,\tau)=\delta(\tau-\tau_{MF})P_a(t),
\end{equation}
where $P_a(t)$ is the probability that a linker is attached regardless of its attachment duration. Using equations~\eqref{eq:PaNormalisation},  \ref{attachment time approximation}), and (\ref{attachment probability approximation}) in (\ref{time evolution of attachment probability}) the rate of change of the probability $P_a(t)$, that a linker be attached at time $t$, is given by the expression,
\begin{equation}
\label{mean field attachment probability evolution}
\partial_{t}P_{a}(t)= \omega_a\left[ 1-P_a(t)\right]-\tau_{MF}^{-1}P_a(t).
\end{equation}
The probability $ P_a$ can be approximated by its steady state value
\begin{equation}
\label{binding probability}
P_{a}^{(s)}(\tau_{MF})=\frac{\omega_a}{\omega_a+\tau^{-1}_{MF}},
\end{equation}
which is obtained by setting the left hand side of equation (\ref{mean field attachment probability evolution}) to zero. Using expressions (\ref{association probability}) and (\ref{binding probability}) in (\ref{cross linking probability}) the cross-linking probability becomes
\begin{equation}
P_{\text{x-link}}=\frac{\delta^3}{R_0^3}\frac{\omega_a }{\omega_a+\tau^{-1}_{MF}}.
\label{eq:Px}
\end{equation}
The number of associated cross-linkers on the reference chain at steady state in expression (\ref{number of associated linkers}) can therefore be approximated as 
\begin{equation}
\label{linear cross-linker number}
N_x\approx\left(\frac{M}{M_e}\right)^2\frac{\delta^3}{R_0^3}\frac{\omega_a f^2}{\omega_a+\tau^{-1}_{MF}}=\left(\frac{N}{N_e}\right)^{\frac{1}{2}}\frac{\omega_a f^2}{\omega_a+\tau^{-1}_{MF}}.
\end{equation}
This comes from considering the random walk nature of the chains which suggests that $\delta= N_e^{1/2}b$ and  $R_0= N^{1/2}b$. To calculate the reversible cross-link induced force per unit velocity of the center of mass. First, we need an expression for attached linker liker tension. This can be written in terms of the velocity of the center of mass of the primitive chain. 

\subsection{Linker friction approximation}

In the limit of short attachment times, the tether at $s_0$ that has been associated for a time $\tau$, has an extension
\begin{equation}
\label{extension-velocity exp}
\bm{\Delta}(t,\tau;s_0)\approx \bm{r}(t,s_0)- \bm{r}(t-\tau,s_0)=\tau \bm{v}(s_0,t)
\end{equation}
in time $t$, in response to a deformation.
This leads to a linker associated average tension force on the chain during time interval $\Delta\tau\approx \tau_{MF}$,

\begin{eqnarray}
\nonumber
\label{everage tension force during}
\bm{\Gamma}_L(\bm{v})= \frac{1}{ \tau_\text{MF}}N_{\text{x}}\kappa \bm{v} \int_{0}^{\tau_\text{MF}} d\tau  \,\, \tau= \frac{1}{2}N_{\text{x}}\kappa \bm{v} \tau_{\text{MF}}.\\
\end{eqnarray}
\\
We need an explicit expressions for $\tau_{\text{MF}}$. Using expressions (\ref{detachment rate}) and (\ref{extension-velocity exp}), the $\tau_{\text{MF}}$ is expressed in terms of the velocity as
\begin{equation}
\label{mean field survival time}
\tilde{\tau}_{MF}^{-1}= 1+ \frac{\alpha^2 \bm{v}^2}{\tilde{\tau}_{MF}^{-2}\omega_0^2},
\end{equation}
where $\tilde{\tau}_{MF}=\omega_0\tau_{MF}$. To keep the notation simple we shall drop the label $MF$ in subsequent expressions. The only physical solution of (\ref{mean field survival time}) leads to
\begin{equation}
\label{mean field time simplified}
\tau=\frac{1}{\omega_0}\left(1-\frac{\bm{v}^2\alpha^2}{\omega_0^2}\right).
\end{equation}

Using expression (\ref{linear cross-linker number}) and (\ref{mean field time simplified}) in (\ref{everage tension force during}) leads to the average tension force
\begin{eqnarray}
\nonumber
\label{tension}
\bm{\Gamma}_L(\bm{v})\approx\frac{\kappa \bm{v} N^{\frac{1}{2}}}{\omega_0 N_e^{\frac{1}{2}}} \left(1-\frac{\bm{v}^2\alpha^2}{\omega_0^2}\right)\frac{  \omega_a f^2}{\omega_a+ \omega_0}.\\
\end{eqnarray}
To approximate the reversible cross-link friction on the entangled chain evolution, we expand the tension expression about zero velocity as in equation (\ref{tension expansion}) up to first order. The zeroth order term of the expansion evaluates to $\bm{\Gamma}_L(\bm{v}=0)=0$.
 The linker reaction induced friction coefficient $\zeta_L$ becomes
\begin{equation}
\label{friction coefficient}
\zeta_L \approx \left(\frac{\partial  \bm{\Gamma}_L}{\partial \bm{v}}\right)\left|_{\bm{v}=0}\right. =  \frac{\kappa f^2 N^{\frac{1}{2}}}{\omega_0 N_e^{\frac{1}{2}}}\left( \frac{\omega_a}{\omega_0 + \omega_a}\right).
\end{equation}

This defines the friction coefficient that a point on the primitive chain experiences in terms of the  attachment probability and the linker characteristic attachment and detachment rates $\omega_a$ and $\omega_0$ respectively. The expression (\ref{friction coefficient}) captures the fact that the primitive chain evolution depends on the connecting tether stiffness through the parameter $\kappa$.

\section{Influence of reversible cross-linkers on chain motion}
\label{primitive chain motion}

The weak association assumption restricts the survival time of the reversible linkers $\tau_s$ in the interval $[\tau_e,\tau_R]$. In time $t$ which lies within the interval, if all reversible linker are dissociated the primitive path can move a curvilinear distance distance
\begin{equation}
x^2(t)\approx b^2N\left(t/Z^2t_R^{(0)}\right)^{1/2}\text{ in } \tau_e \leq t \leq \tau_R (N_x).
\end{equation}
When the equilibrium number of linkers are in an associated state, the primitive chain displacement is influenced by anchoring effect of reversible linkers. The chain sub-units that are directly cross-linked can explore space through tether extension.  In this way reversible cross-linking contribute a friction type slowing down effect. The curvilinear distance that can be moved by the primitive chain segments is therefore
\begin{equation}
x^2(t)\approx \frac{b^2N t^{1/2}}{\left(Z^2t_R^{(0)}[1+\frac{\zeta_L}{\zeta_f}]\right)^{1/2}}\text{ in }\tau_e \leq t \leq \tau_R (N_x).
\end{equation}
The new entanglement length due to the additional constraints brought about by the tether is of the order $a\approx a_0\left(\frac{1}{(N_e/N)N_x+1}\right)$ 
where $a_0$ is the entanglement length for the system when reversible linker are switched off. Thus
\begin{equation}
\tau_e\approx \tau_e^{(0)}\left(\frac{1}{(N_e/N)N_x +1}\right)^4 \text{ and } \tau_d\approx Z\tau_R.
\end{equation}
 The Rouse time for the case of weakly associating reversibly cross-linked system is approximated to 
\begin{eqnarray}
\nonumber
 \tau_R &\approx& \tau_R^{(0)}\left(1 +\frac{\zeta_L}{\zeta_f}\right)\\
 \label{modified Rouse time}
 &=& \tau_R^{(0)}\left(1 + \frac{\kappa f^2 N^{\frac{1}{2}}}{\zeta_f\omega_0 N_e^{\frac{1}{2}}}\frac{\omega_a}{\omega_0 + \omega_a}\right).
\end{eqnarray}
where the Rouse time for the system without reversible cross-links is given by $\tau_R^{(0)}=\frac{ N^2b^2\zeta}{3k_{\text{B}}T}$ with $b$ being the bond vector of the chain \cite{doiandEdwards1988}. In times $t \ll\tau_e$ the chain motion is not influenced by both reversibly cross-links and entanglements constraints. The condition for a weak reversible cross-link is met when $\frac{\zeta_L}{\zeta_L+\zeta_f}\ll 1$. Here $\zeta_f\approx N \zeta$ where $\zeta$ is the characteristic monomeric friction.


A connection between stress relaxation is made by imagining that an small external deformation is applied to the system at time $t=0$ and the system is allowed to relax such that the chains restore their equilibrium length. As this happens both entanglements and reversible  cross-links behave like momentary cross-links that can sustain stress as long as they are not released for the first time. For polymer fluids it is well known \cite{doiandEdwards1988,doi2013} that, under shear strain $\gamma$, stress is given by
\begin{equation}
\sigma_{xy}= G_0 \gamma,
\end{equation}
where $G_0$ is the shear modulus. In a reversible entangled network shear strain is proportional to the fraction of constraints that are deformed and have not been released in time $t$
\begin{equation}
\sigma_{xy}(t)= G_0 \gamma\psi(t).
\end{equation}
The relaxation is attributed to disengagement of the constraints. This happens  in time $t> t_e$, where $t_e$ is the entanglement time. This leads to the time dependent relaxation modulus
\begin{equation}
\label{relaxation modulus}
G(t)= G_0 \psi(t).
\end{equation}

As highlighted earlier, in the studied system, any given test chain that relaxes has two types of temporary constraints: the entanglements and the reversible cross-linkers. In the expressions above, note that we need to find the timescales and mechanisms for the relaxation of these constraints.  The chain can diffuse out of the entanglements so that these become released and the reversible cross-linkers are released via another process, if we assume that such a reversible cross-linker does not diffuse along the chain. In a preceding section we modelled the release process, which is in part simply dependent on a rate and also on the local stress applied to the linker.  Clearly, the two release processes are coupled. Here we approximate simply that the diffusion coefficient is much reduced due to cross-linker friction as  motivated earlier. We ignore possible effects at the ends where the two effects do influence each other.

\begin{figure}[ht]
 \includegraphics[width=.5\textwidth, angle=0]{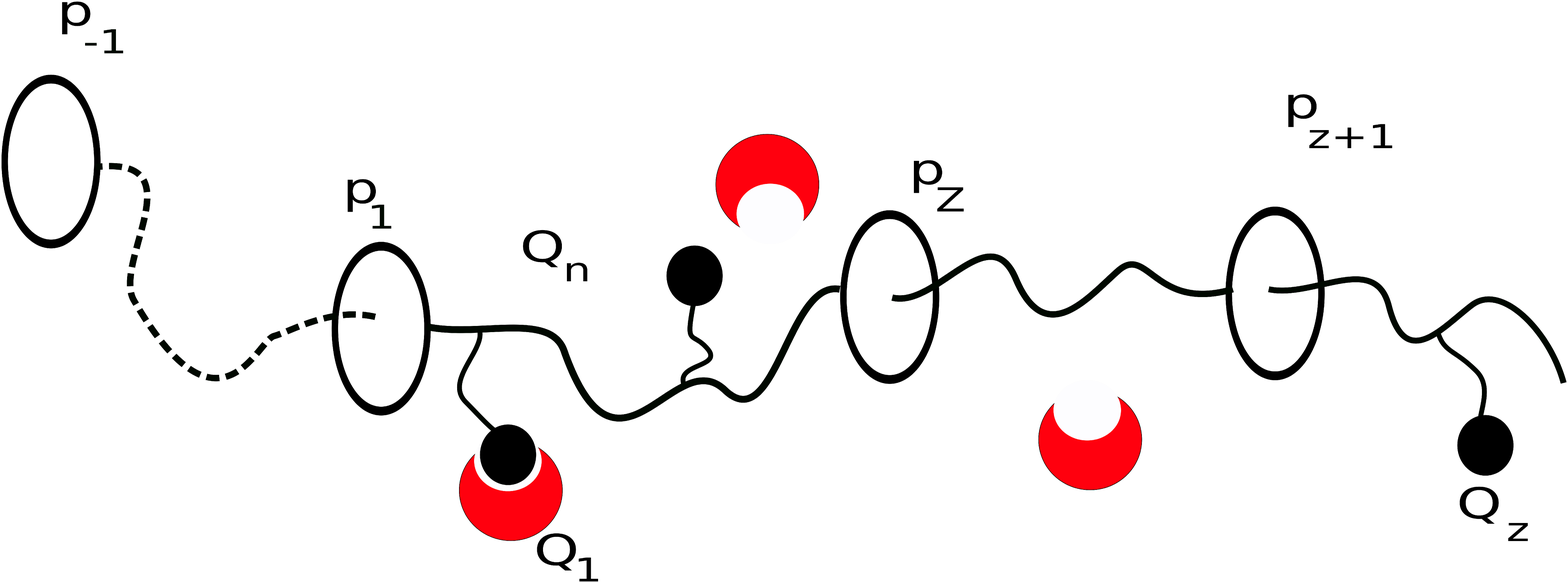}
	\caption{The process of creation and destruction of constraints in reversible entangled network.}	\label{constraints formation and release}
\end{figure}

As discussed earlier, in a typical slip-link model, entanglement junctions are presented by small rings. This does not suggest that the actual constraint due to entanglements is a localised type of topological constraint. The assumption entails the notion that as two polymers get entangled, their evolution on a spatial length of the order of the distance between slip-links get coupled for a certain period of time which is of the order of the time between creation and distraction
of slip-links. Like-wise reversible cross-links association reactions lead to the constraining of the motion of the two cross-linked parts of the chains, for a period of the order of the cross-link survival time. This happens over a certain spatial length which is presumably determined by the length of the tethers or spacing between entanglements. 
Let us suppose the chain in FIG. \ref{constraints formation and release} moves to the right. The reversible linker $Q_n$ or a slip-link $P_n$ are destroyed. As this happens a new slip-link $P_{Z+1}$ or a new cross-link $Q_{Z+1}$ may be created.

In a situation where a small deformation is applied to the system, the chains are deformed generating stress internally. As the chains relax to their equilibrium conformations stress is released. As stated earlier the reptation process is slowed down owing to the fact that chain sub-units have to stretch tethers that are bound by sticky groups.
We have to introduce some assumptions to approximate the relaxation modulus. Let us assume that reversible cross-links tethers can sustain an externally applied stress only if they have not detached for this first time after an application of stress. In their relaxed equilibrium  extension reversible linker contribute friction to the diffusion of the chain between entanglements.
The relaxation modulus can therefore be easily approximated with if we know the fraction $\psi(t)$ of constraints that still sustain stress in time $t$.

\section{Releasing constraints after strain: entanglements and reversible linkers}
\label{sec:release}

We consider here the processes that occur when the system relaxes after a strain. The fundamental relaxation process for a melt of polymer chains will be reptation, with the release of constraints, occurring as the chain diffuses out of its tube, or equivalent set of discrete entanglement constraints. Doi \cite{doi2013} presents a treatment of the discrete entanglement constraints. Variations of the relaxation due to other related effects that include additional slow time scales refine this treatment\cite{MilnerandMcLeish1998,LikhtmanandMcLeish2002}.

In our considerations we presume the reversible cross-linkers attach and detach far more rapidly than the slow diffusive reptation-related processes. This is in stark contrast to the cases considered by Rubinstein and Semenov\cite{RubinsteinANDSemenov1998}. In this latter scenario the reversible cross-linkers act as ``gate keepers'' to the continuation of the reptation process. 

In the previous sections we interpreted one of the roles of rapid attachment and detachment as leading to an effective enhanced friction. Here, following a step-strain, it is useful to consider how the faster reversible cross-linking process releases its own constraints, and is coupled to the slower motion of the tube.

The chain now has two different types of constraints, \emph{viz.}~the typical entanglements constraints, but also the constraints due to the reversible linking. The process of reptation cause the release of the entanglements constraints.
Whereas the chain must diffuse for the entanglements to be released, the reversible cross-links detachment is dependent primarily on the longevity of the reversible bond, and also on the stress applied to the reversible bond. Even though a bond may rapidly re-attach to the chain, before re-attaching at a later stage some of the stress of the chain will be released subsequent to the reversible bond detachment. Although the process is certainly subtle, we estimate coarse time scales here.

We recall that for the detachment process the number of surviving, of the original $N_x$ initially attached linkers, is $N_\text{remain}(t) = N_x \exp \left( - \omega_0 t \right)$. We ignore, for simplicity, the stretch dependent correction. Given an arc length on average between entanglements, and an initial average density of attached reversible linkers (related to eq.~\eqref{eq:Px}) we know how many reversible linkers must also be released per released entanglement.

We now consider two constraint relaxation processes in addition to the reptation of the chain in its tube.

An initial process after strain (before the release of entanglements) entails the longitudinal relaxation of the segments between entanglements \cite{MilnerandMcLeish1998,LikhtmanandMcLeish2002}. Whereas without temporary cross-links this relaxation is given by Rouse behaviour up to the length scale $a$ between entanglements. This relaxation happen within Rouse time scale $\tau_{R}^{(0)}$. We expect again that, with reversible cross-links, this process is retarded by detachments and re-attachments as before hence with an additional friction. However, since the chain has, after such a short time not released entanglements, this can only accommodate a small amount of relaxation. In the extreme case, with very low re-attachment rates one can argue that when $N_\text{remain} <1$, all attached chain sections becomes released leading on average to a time-scale $\tau_{R}(N_x)$ over which a chain sub-unit that was
associated with a reversible cross-link is relaxed. 

The second such process entails the chain ends that are released from an entanglement constraint. Thinking in the picture of Doi\cite{DoiandEdwards1978} slip-link constraints are released as a chain diffuses out of the link, freeing an end of arc length $a$, each time a single entanglement constraint is released. With reversible cross-links the freeing from the slipping link, allows the relaxation of the reversibly attached linkers on this length of chain. There are approximately $N_\text{x}\frac{a}{L}$ such linkers to be released, the release of each in sequence away from the free end takes an average time $\tau_s=1/\omega_0$.

The fraction of the chain that has not fully relaxed stress is proportional to the sum of the three contributions above:
\begin{equation}
\label{fraction of unrelaxed constraints}
	\psi(t)= \psi_{\text{init}}(t) + \psi_{\text{ent}}(t) + \psi_{\text{end}}(t),
\end{equation}
with $\psi_\text{init}$ approximates 
the low frequency longitudinal Rouse motion on time scales longer than $\tau_e$ \cite{MilnerandMcLeish1998,LikhtmanandMcLeish2002} but modified by the additional friction due to reversible linkers. The term $\psi_{ent}$ is the contribution from entanglements while the term $\psi_{\text{end}}(t)$ is the contribution from the ends released from the entanglement, but still constrained by the reversible linkers for a short time $\tau_s$.

\subsection{Initial release process}

The complete longitudinal relaxation of the chain segments in between entanglements happen at different times depending on whether a segment is directly associated with a reversible cross-link or not. About $N_x$ of the chain subunits bearing  the attached constraints will have a retarded relaxation. Thus we can split the longitudinal relaxation into two contributions 
\begin{eqnarray}
\nonumber
\psi_\text{init}(t)&=&\frac{1}{3 Z}\sum_{p=Z+1}^{Z+N_x}\exp(-p^2 t/\tau_{R})
\label{initial unreleased constraints fraction}
\\&+&
 \frac{1}{3 Z}\sum_{p=1}^{Z}\exp(-p^2 t/\tau_{R}^{(0)}) .
\end{eqnarray}
Here $\tau_{R}(N_x)$ is the reversible linker delayed Rouse relaxation time approximate in (\ref{modified Rouse time}) and $\tau_{R}{(0)}$ is the usual Rouse time of a chain without Reversible cross-links. The factor of three comes from the fact that only one of the three vector components of the Rouse modes is unhindered by the tube. The first term captures the initial (partial) relaxation of longitudinal Rouse modes of the chain which are delayed by the reversible cross-linker attachment while the second term captures the low frequency contribution for sub-units in between entanglements
that are not directly associated with reversible linkers. Reversible linkers may reattach to the part of the chain in between slip-links. A reattachment to part of the chain which has gone through longitudinal relaxation only influence the long time diffusive behavior of the system through friction. 

\subsection{Entanglement release process}

The well known fraction of the chain that is constrained by the original slip-links is given by
\begin{equation}
\label{fraction of unreleased slip-links}
\psi_\text{ent}(t)= \frac{1}{Z}\sum_\text{n=1}^{Z}\int_{0}^{L}
ds\psi_{ent}^{(n)}(s,t).
\end{equation}
The function $\psi^{(n)}(s,t)$ may also be interpreted as the probability that the entanglement point $P_n$ remains unreleased at a time $t$ while the right end of the chain at is at $s$. This has been calculated in detail for the tube model \cite{doi2013} and it is known to satisfy the following diffusion equation 
\begin{equation}
	\partial_t \psi_\text{ent}^{(n)}(t,s) = D_c \partial^2_s \psi_\text{ent}^{(n)}(t,s),
\end{equation}
under the boundary condition
 \begin{equation}\psi_{\text{ent}}^{(n)}(0,t)=0\text{ and } \psi_{\text{ent}}^{(n)}(L,t)=0.
 \end{equation}
and initial condition
 \begin{equation}\psi_{\text{ent}}^{(n)}(s,0)=\delta(s-[L-n a]),
 \end{equation}
leading to the solution 
\begin{eqnarray} \lefteqn{\psi_{\text{ent}}^{(n)}(s,t)}
	\nonumber \\
& =&
\nonumber
	\frac{1}{L}\left(\sum_{p=1}^{\infty} \sin{\left(\frac{p \pi s}{L}\right)}\sin{\left(\frac{p\pi (L-n a_x)}{L}\right)}
 \exp{\left[-\frac{ t}{\tau_d }p^2\right]}\right)\\
\end{eqnarray}
where 
\begin{equation}
\tau_d=\frac{L^2}{\pi^2 D_c}
\end{equation}
is the life time of an entanglement.

Converting the sum over $n$ into an integral in expression (\ref{fraction of unreleased slip-links}), the fraction of the primitive chain constrained by the original slip-links is given by
\begin{eqnarray}
\label{slip-links fraction}
\nonumber	\psi_{\text{ent}}(t) &=&  \int d n\int d s\, \psi_\text{ent}^{(n)}(t,s)\\
\label{original slip-links fraction}
&=&\sum_{p:\text{odd}}^{\infty}
\frac{8}{\pi^2p^2}\exp(-t/\tau_d)
\end{eqnarray}

\subsection{Release of reversible linkers following release of end entanglement}

For the conventional slip-link model a fraction of the chain that is released from the original tube is assumed to relax stress almost instantaneously. With reversible cross-linking at short time scales we need to include the additional detachment process of linkers on that part of the chain. This means that the fast retraction motion of the chain ends is delayed. So, the entanglement's release leads to a sequence of reversible bond release processes of the free end. Re-attachments then can happen to the relaxed chain and are not counted in the same way as new entanglements are not counted for the strain response.

\begin{widetext}
The average number of released end constraints depends on when the constraints were released and on how much time has elapsed since their release. The rate at which entanglements are released at any time is the derivative of $\psi_\text{ent}(t)$, and again we have the decay of the attached linkers as described above:
\begin{align}
\nonumber
	\psi_{\text{end}} (t) &=-P_{\text{x-link}} \int_0^t \mathrm{d}\tau \, \frac{\mathrm{d} \psi_{\text{ent}}(\tau) }{\mathrm{d} \tau}
	 e^{- (t-\tau)/\tau_s} \\
\label{partially relaxed chain end}
&=P_{\text{x-link}}\sum_{p:\text{odd}}^{\infty}\frac{8}{\pi p^2}\left[\left( 1- \frac{\tau_d}{\tau_d-p^2\tau_s} \right)e^{-t/\tau_s}+\left(  \frac{\tau_d}{\tau_d-p^2\tau_s} -1\right)e^{-tp^2/\tau_d}\right].
\end{align}
Here $\psi_{\text{ent}}$ is as given in (\ref{slip-links fraction}). The expression above pre-supposes that the reversible cross-linkers can relax in any sequence. This approximation is reasonable if entanglements are released slowly and reversible cross-linkers rapidly thereafter. 

Upon using (\ref{initial unreleased constraints fraction}),(\ref{original slip-links fraction}) and (\ref{partially relaxed chain end}) in (\ref{fraction of unrelaxed constraints}), the sum of the fraction of unrelated constraints from the three contributions becomes
\begin{eqnarray}
\nonumber
\psi(t)&=& \frac{1}{3}\sum_{p=Z}^{Z+N_x}e^{-p^2 t/\tau_{R}} +\frac{1}{3}\sum_{p=Z+N_x}^{N}e^{-p^2 t/\tau_{R}^{(0)}}
\\
&+& P_{\text{x-link}}\sum_{p:\text{odd}}^{\infty}\frac{8}{\pi p^2}\left( 1- \frac{\tau_d}{\tau_d-p^2\tau_s} \right)e^{-t/\tau_s}+\sum_{p:\text{odd}}^{\infty}\frac{8}{\pi p^2}\left[1+P_{\text{x-link}}\left(  \frac{\tau_d}{\tau_d-p^2\tau_s} -1\right)e^{-tp^2/\tau_d}\right]
\end{eqnarray}

\end{widetext}

\section{Stress Relaxation}
\label{Stress Relaxation}

In contrast to non-sticky entangled chain which has a modulus that decays with a single relaxation time, we learn from (\ref{relaxation modulus})  that the modulus of a reversibly cross-linked entangled chain declines as a sum of two exponential relaxation curves each with  relaxation times $\tau_d$ and $\tau_s$.
The comparison becomes even more clear on looking at the linear viscoelastic theoretical predictions for an entangled system with its reversibly cross-linked entangled counterpart  as characterised by the storage modulus
\begin{eqnarray}
\nonumber
G'(\omega)&=& \omega\int_0^{\infty} \sin(\omega t) G(t)\\
\nonumber
&=&\frac{1}{3}\left[\sum_{p=Z}^{Z+N_x} \frac{G_N^{(0)}(\omega \tau_{R} /p^2)}{1+((\omega \tau_{R})/p^2)^2}+\sum_{p=Z+N_x}^{N} \frac{G_N^{(0)}(\omega \tau_{R}^{(0)} /p^2)}{1+((\omega \tau_{R}^{(0)})/p^2)^2}\right]\\
&+&\nonumber G_N^{(0)}P_{\text{x-link}}\sum_{p:\text{odd}}^{\infty}\frac{8}{\pi p^2}\left( 1- \frac{\tau_d}{\tau_d-p^2\tau_s} \right)\frac{(\omega \tau_s /p^2)}{1+((\omega \tau_s)/p^2)^2}\\
\nonumber
&+&
\label{storage modulus}
G_N^{(0)}\sum_{p:\text{odd}}\frac{8}{(\pi p)^2}\left[1+P_{\text{x-link}}\left(  \frac{\tau_d}{\tau_d-p^2\tau_s} -1\right)\right]\\
&\times&\frac{(\omega \tau_d /p^2)}{1+((\omega \tau_d)/p^2)^2}.
\end{eqnarray}
Equation (\ref{storage modulus}) holds for times $t\geq\tau_e$. At times $t<<\tau_e$ the Rouse segments motion is almost independent of the constraints. The elastic modulus is  given by 
\begin{equation}
\label{storage modulus at short times}
G'(\omega)=G_N^{(0)}\left(\frac{\pi}{2}\omega \tau_e^{(0)}\right)^{1/2}.
\end{equation}
For a detailed derivation of the expression (\ref{storage modulus at short times}) see \cite{doiandEdwards1988}.

\begin{figure}[ht]
 \includegraphics[width=.48\textwidth, angle=0]{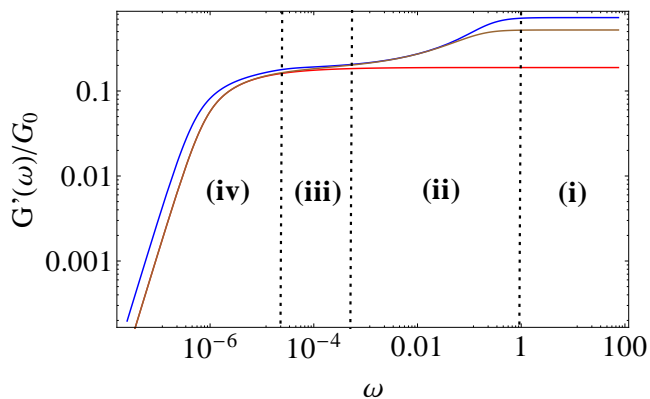}
	\caption{Plot of the storage modulus according to models including different effects for $\{N=500, N_e=4, \tau_s =2.0,\zeta=1.0, f=0.1,\omega_a=0.50, \omega_0= 0.50, k=3.0, b=1.0, k_\textbf{B}T=1.0\}$. The blue curve reflects the full behaviour with the reversible cross-linker effects dicussed in this paper, the red simple reptation, and the brown reptation with inter-entanglement relaxation. More details, and a description of the depicted regimes (i)--(iv) are provided in the main text. }\label{graph of storage modulus}
\end{figure}

FIG.~\ref{graph of storage modulus} shows the storage modulus in equation (\ref{storage modulus}) normalized by the plateau value. The three curves portray the storage modulus when different contributions are are taken into account. The red line shows the relative storage modulus when the relaxation mechanism is assumed to be the diffusive motion out of slip-links with no cross-reversible linker effects included, and serves as a reference line. The brown curve depicts the relative storage modulus when longitudinal relaxation of subunits between slip-links is assumed but still with no reversible linkers\cite{MilnerandMcLeish1998,LikhtmanandMcLeish2002}. This is the case when $P_{\text{x-link}} =0$ in equation (\ref{storage modulus}) while the blue line is the graph of the relative storage modulus for an entangled reversibly cross-linked system that relax through reptation and longitudinal relaxation. In this case the influence of reversible linkers is taken into account.

As illustrated in FIG.~\ref{graph of storage modulus}, the derived expression for the elastic modulus captures several regimes of relaxation behavior marked (i) to (iv) and also portrays the role of weakly associating reversible linkers:
\begin{itemize}
	\item {\textbf{(i)--(ii):}}  We see the highest frequency plateau where chain subunits between linkers and reversible linker tethers relax. In this case the reversible linker contribution leads to an increase in plateau plateau height in comparison to the case when they are switched off (\emph{i.e.} the brown line which includes segmental relaxation.) We note that the segmental relaxation is again delayed by the effective friction (with short-lived reversible linking).
	The rise of a plateau in zone (i) from the plateau in zone (ii) occurs when chain subunits between entanglements undergo transverse relaxation
	\item {\textbf{(iii)--(iv):}} The main plateau due to entanglements. On this time scale reversible cross-links have already detached in the processes in regions (i)--(ii). At very long time scales (iv), the effect of the additional effective drag produced by  reversible cross-linking is noticeable.
\end{itemize}

\section{Discussion}
\label{Discussion}

In this paper we discuss the role of weakly associating reversible cross-linkers on the diffusion of a primitive chain through slip-links.
By approximating the slowing down effect of the associated groups, on the chain motion, as an effective friction, the contribution of the associating groups to the relaxation modulus is calculated.

From the relaxation modulus we learn that in time scales $\tau_e(N_{x})\leq t \leq\tau_R(N_x)$ the storage modulus $G'(t)$ has a higher second plateau compared to the case when reversible linkers are switched off.
This is due to contributions from reversible cross-links on the elastic behaviour of the system. This contribution is in addition to a permanently cross-linked network like plateau which emanates from the entanglements. The reversible cross-linkers also lead to a delayed longest relaxation time in contrast to a purely entangled system. We therefore argue that despite the degree of movement presented by the tethers, at time scales $t<\tau_s$, a weakly associating melt will look comparable to a permanently cross-linked network. However, in times $t>\tau_s$ the system will look analogous to an entangled melt system.

In the case where the reversible linkers survival time is greater Rouse time  $\tau_s > \tau_R(N_x)$, the steric effects get more pronounced leading to a localisation of cross-link bearing segments within the network mersh size. An entangled system with tethered reversibly cross-linkers proves to have a  modulus analogous that predicted for sticky reptation model \cite{leibler1991}. The predicted storage modulus, even in this case of cross-linkers with of short survival time, has four distinct regimes that are separated by time scales $\tau_e, \tau_R$ and $\tau_d$.  

From the qualitative results obtained we can infer that the storage modulus additional plateau observed in the experimental data of Das.et.al. \citep{Das2015} possibly originates from slowing down effect related to the associating groups. We do not attempt any comparison of our qualitative prediction with experimental data, Such comparison require further details on the kinetic parameters and the spring constant of the connecting tethers, which calls for further experimental investigations.

For not so long entangled chains with weakly associating groups, many-chain relaxation mechanisms such as primitive path length fluctuations and dynamic constraint release are expected to play a non-negligible role. Further extension of the discussion on the role of the weakly associating groups will take such relaxation mechanisms into account.
 
\textbf{Aknowledgement.} The authors specifically aknowledge
A. Das and K.W St\"{o}ckelhuber for arousing our interest in studying entangled reversible networks dynamics. M.J. Mateyisi
is grateful to J. Peturej, T. Kreer, L. Ivaneiko and M. Kock for stimulating discussions.

\bibliography{MJM_GH_kkMN_preprint.bib}

\begin{thebibliography}{21}%
\makeatletter
\providecommand \@ifxundefined [1]{%
 \@ifx{#1\undefined}
}%
\providecommand \@ifnum [1]{%
 \ifnum #1\expandafter \@firstoftwo
 \else \expandafter \@secondoftwo
 \fi
}%
\providecommand \@ifx [1]{%
 \ifx #1\expandafter \@firstoftwo
 \else \expandafter \@secondoftwo
 \fi
}%
\providecommand \natexlab [1]{#1}%
\providecommand \enquote  [1]{``#1''}%
\providecommand \bibnamefont  [1]{#1}%
\providecommand \bibfnamefont [1]{#1}%
\providecommand \citenamefont [1]{#1}%
\providecommand \href@noop [0]{\@secondoftwo}%
\providecommand \href [0]{\begingroup \@sanitize@url \@href}%
\providecommand \@href[1]{\@@startlink{#1}\@@href}%
\providecommand \@@href[1]{\endgroup#1\@@endlink}%
\providecommand \@sanitize@url [0]{\catcode `\\12\catcode `\$12\catcode
  `\&12\catcode `\#12\catcode `\^12\catcode `\_12\catcode `\%12\relax}%
\providecommand \@@startlink[1]{}%
\providecommand \@@endlink[0]{}%
\providecommand \url  [0]{\begingroup\@sanitize@url \@url }%
\providecommand \@url [1]{\endgroup\@href {#1}{\urlprefix }}%
\providecommand \urlprefix  [0]{URL }%
\providecommand \Eprint [0]{\href }%
\providecommand \doibase [0]{http://dx.doi.org/}%
\providecommand \selectlanguage [0]{\@gobble}%
\providecommand \bibinfo  [0]{\@secondoftwo}%
\providecommand \bibfield  [0]{\@secondoftwo}%
\providecommand \translation [1]{[#1]}%
\providecommand \BibitemOpen [0]{}%
\providecommand \bibitemStop [0]{}%
\providecommand \bibitemNoStop [0]{.\EOS\space}%
\providecommand \EOS [0]{\spacefactor3000\relax}%
\providecommand \BibitemShut  [1]{\csname bibitem#1\endcsname}%
\let\auto@bib@innerbib\@empty
\bibitem [{\citenamefont {Leibler}, \citenamefont {Rubinstein},\ and\
  \citenamefont {Colby}(1991)}]{leibler1991}%
  \BibitemOpen
  \bibfield  {author} {\bibinfo {author} {\bibfnamefont {L.}~\bibnamefont
  {Leibler}}, \bibinfo {author} {\bibfnamefont {M.}~\bibnamefont {Rubinstein}},
  \ and\ \bibinfo {author} {\bibfnamefont {R.~H.}\ \bibnamefont {Colby}},\
  }\bibfield  {title} {\enquote {\bibinfo {title} {Dynamics of reversible
  networks},}\ }\href {\doibase 10.1021/ma00016a034} {\bibfield  {journal}
  {\bibinfo  {journal} {Macromolecules}\ }\textbf {\bibinfo {volume} {24}},\
  \bibinfo {pages} {4701--4707} (\bibinfo {year} {1991})},\ \Eprint
  {http://arxiv.org/abs/http://dx.doi.org/10.1021/ma00016a034}
  {http://dx.doi.org/10.1021/ma00016a034} \BibitemShut {NoStop}%
\bibitem [{\citenamefont {Tanaka}\ and\ \citenamefont
  {Edwards}(1992)}]{tanaka1992viscoelastic}%
  \BibitemOpen
  \bibfield  {author} {\bibinfo {author} {\bibfnamefont {F.}~\bibnamefont
  {Tanaka}}\ and\ \bibinfo {author} {\bibfnamefont {S.}~\bibnamefont
  {Edwards}},\ }\bibfield  {title} {\enquote {\bibinfo {title} {Viscoelastic
  properties of physically crosslinked networks: Part 1. non-linear stationary
  viscoelasticity},}\ }\href@noop {} {\bibfield  {journal} {\bibinfo  {journal}
  {Journal of Non-Newtonian Fluid Mechanics}\ }\textbf {\bibinfo {volume}
  {43}},\ \bibinfo {pages} {247--271} (\bibinfo {year} {1992})}\BibitemShut
  {NoStop}%
\bibitem [{\citenamefont {Das}\ \emph {et~al.}(2015)\citenamefont {Das},
  \citenamefont {Sallat}, \citenamefont {Böhme}, \citenamefont {Suckow},
  \citenamefont {Basu}, \citenamefont {Wießner}, \citenamefont
  {Stöckelhuber}, \citenamefont {Voit},\ and\ \citenamefont
  {Heinrich}}]{Das2015}%
  \BibitemOpen
  \bibfield  {author} {\bibinfo {author} {\bibfnamefont {A.}~\bibnamefont
  {Das}}, \bibinfo {author} {\bibfnamefont {A.}~\bibnamefont {Sallat}},
  \bibinfo {author} {\bibfnamefont {F.}~\bibnamefont {Böhme}}, \bibinfo
  {author} {\bibfnamefont {M.}~\bibnamefont {Suckow}}, \bibinfo {author}
  {\bibfnamefont {D.}~\bibnamefont {Basu}}, \bibinfo {author} {\bibfnamefont
  {S.}~\bibnamefont {Wießner}}, \bibinfo {author} {\bibfnamefont {K.~W.}\
  \bibnamefont {Stöckelhuber}}, \bibinfo {author} {\bibfnamefont
  {B.}~\bibnamefont {Voit}}, \ and\ \bibinfo {author} {\bibfnamefont
  {G.}~\bibnamefont {Heinrich}},\ }\bibfield  {title} {\enquote {\bibinfo
  {title} {Ionic modification turns commercial rubber into a self-healing
  material},}\ }\href {\doibase 10.1021/acsami.5b05041} {\bibfield  {journal}
  {\bibinfo  {journal} {ACS Applied Materials \& Interfaces}\ }\textbf
  {\bibinfo {volume} {7}},\ \bibinfo {pages} {20623--20630} (\bibinfo {year}
  {2015})},\ \bibinfo {note} {pMID: 26332010},\ \Eprint
  {http://arxiv.org/abs/http://dx.doi.org/10.1021/acsami.5b05041}
  {http://dx.doi.org/10.1021/acsami.5b05041} \BibitemShut {NoStop}%
\bibitem [{\citenamefont {TCates}(1988)}]{Cates1988}%
  \BibitemOpen
  \bibfield  {author} {\bibinfo {author} {\bibfnamefont {M.~E.}\ \bibnamefont
  {TCates}},\ }\bibfield  {title} {\enquote {\bibinfo {title} {Stress
  relaxation and chemical kinetics in pairwise associating polymers},}\ }\href
  {\doibase 10.1021/ma00179a049} {\bibfield  {journal} {\bibinfo  {journal}
  {Macromolecules}\ }\textbf {\bibinfo {volume} {21}},\ \bibinfo {pages}
  {256--259} (\bibinfo {year} {1988})}\BibitemShut {NoStop}%
\bibitem [{\citenamefont {Rubinstein}\ and\ \citenamefont
  {Semenov}(2001)}]{RubinsteinANDSemenov2001}%
  \BibitemOpen
  \bibfield  {author} {\bibinfo {author} {\bibfnamefont {M.}~\bibnamefont
  {Rubinstein}}\ and\ \bibinfo {author} {\bibfnamefont {A.~N.}\ \bibnamefont
  {Semenov}},\ }\bibfield  {title} {\enquote {\bibinfo {title} {Dynamics of
  entangled solutions of associating polymers},}\ }\href {\doibase
  10.1021/ma0013049} {\bibfield  {journal} {\bibinfo  {journal}
  {Macromolecules}\ }\textbf {\bibinfo {volume} {34}},\ \bibinfo {pages}
  {1058--1068} (\bibinfo {year} {2001})}\BibitemShut {NoStop}%
\bibitem [{\citenamefont {Rubinstein}\ and\ \citenamefont
  {Semenov}(1998)}]{RubinsteinANDSemenov1998}%
  \BibitemOpen
  \bibfield  {author} {\bibinfo {author} {\bibfnamefont {M.}~\bibnamefont
  {Rubinstein}}\ and\ \bibinfo {author} {\bibfnamefont {A.~N.}\ \bibnamefont
  {Semenov}},\ }\bibfield  {title} {\enquote {\bibinfo {title}
  {Thermoreversible gelation in solutions of associating polymers. 2. linear
  dynamics},}\ }\href {http://dx.doi.org/10.1021/ma970617+} {\bibfield
  {journal} {\bibinfo  {journal} {Macromolecules}\ }\textbf {\bibinfo {volume}
  {31}},\ \bibinfo {pages} {1386--1397} (\bibinfo {year} {1998})},\ \Eprint
  {http://arxiv.org/abs/http://dx.doi.org/10.1021/ma970617+}
  {http://dx.doi.org/10.1021/ma970617+} \BibitemShut {NoStop}%
\bibitem [{\citenamefont {Semenov}(2006)}]{Semenov2006}%
  \BibitemOpen
  \bibfield  {author} {\bibinfo {author} {\bibfnamefont {A.~N.}\ \bibnamefont
  {Semenov}},\ }\bibfield  {title} {\enquote {\bibinfo {title} {Dynamics of
  associating polymers with random structure},}\ }\href
  {http://stacks.iop.org/0295-5075/76/i=6/a=1116} {\bibfield  {journal}
  {\bibinfo  {journal} {EPL (Europhysics Letters)}\ }\textbf {\bibinfo {volume}
  {76}},\ \bibinfo {pages} {1116} (\bibinfo {year} {2006})}\BibitemShut
  {NoStop}%
\bibitem [{\citenamefont {Ahmadi}\ \emph {et~al.}(2015)\citenamefont {Ahmadi},
  \citenamefont {Hawke}, \citenamefont {Goldansaz},\ and\ \citenamefont {van
  Ruymbeke}}]{vanRuymbeke2015}%
  \BibitemOpen
  \bibfield  {author} {\bibinfo {author} {\bibfnamefont {M.}~\bibnamefont
  {Ahmadi}}, \bibinfo {author} {\bibfnamefont {L.~G.~D.}\ \bibnamefont
  {Hawke}}, \bibinfo {author} {\bibfnamefont {H.}~\bibnamefont {Goldansaz}}, \
  and\ \bibinfo {author} {\bibfnamefont {E.}~\bibnamefont {van Ruymbeke}},\
  }\bibfield  {title} {\enquote {\bibinfo {title} {Dynamics of entangled linear
  supramolecular chains with sticky side groups: Influence of hindered
  fluctuations},}\ }\href {\doibase 10.1021/acs.macromol.5b00733} {\bibfield
  {journal} {\bibinfo  {journal} {Macromolecules}\ }\textbf {\bibinfo {volume}
  {48}},\ \bibinfo {pages} {7300--7310} (\bibinfo {year} {2015})},\ \Eprint
  {http://arxiv.org/abs/http://dx.doi.org/10.1021/acs.macromol.5b00733}
  {http://dx.doi.org/10.1021/acs.macromol.5b00733} \BibitemShut {NoStop}%
\bibitem [{\citenamefont {Hoy}\ and\ \citenamefont
  {Fredrickson}(2009)}]{Fredrickson2009}%
  \BibitemOpen
  \bibfield  {author} {\bibinfo {author} {\bibfnamefont {R.~S.}\ \bibnamefont
  {Hoy}}\ and\ \bibinfo {author} {\bibfnamefont {G.~H.}\ \bibnamefont
  {Fredrickson}},\ }\bibfield  {title} {\enquote {\bibinfo {title}
  {Thermoreversible associating polymer networks. i. interplay of
  thermodynamics, chemical kinetics, and polymer physics},}\ }\href {\doibase
  http://dx.doi.org/10.1063/1.3268777} {\bibfield  {journal} {\bibinfo
  {journal} {The Journal of Chemical Physics}\ }\textbf {\bibinfo {volume}
  {131}},\ \bibinfo {eid} {224902} (\bibinfo {year} {2009}),\
  http://dx.doi.org/10.1063/1.3268777}\BibitemShut {NoStop}%
\bibitem [{\citenamefont {de~Gennes}(1971)}]{deGennes1971}%
  \BibitemOpen
  \bibfield  {author} {\bibinfo {author} {\bibfnamefont {P.~G.}\ \bibnamefont
  {de~Gennes}},\ }\bibfield  {title} {\enquote {\bibinfo {title} {Reptation of
  a polymer chain in the presence of fixed obstacles},}\ }\href {\doibase
  http://dx.doi.org/10.1063/1.1675789} {\bibfield  {journal} {\bibinfo
  {journal} {The Journal of Chemical Physics}\ }\textbf {\bibinfo {volume}
  {55}},\ \bibinfo {pages} {572--579} (\bibinfo {year} {1971})}\BibitemShut
  {NoStop}%
\bibitem [{\citenamefont {Doi}\ and\ \citenamefont
  {Edwards}(1986)}]{doiandEdwards1988}%
  \BibitemOpen
  \bibfield  {author} {\bibinfo {author} {\bibfnamefont {M.}~\bibnamefont
  {Doi}}\ and\ \bibinfo {author} {\bibfnamefont {S.~F.}\ \bibnamefont
  {Edwards}},\ }\href@noop {} {\emph {\bibinfo {title} {The theory of polymer
  dynamics}}}\ (\bibinfo  {publisher} {oxford university press},\ \bibinfo
  {year} {1986})\BibitemShut {NoStop}%
\bibitem [{\citenamefont {Doi}(2013)}]{doi2013}%
  \BibitemOpen
  \bibfield  {author} {\bibinfo {author} {\bibfnamefont {M.}~\bibnamefont
  {Doi}},\ }\href@noop {} {\emph {\bibinfo {title} {Soft matter physics}}}\
  (\bibinfo  {publisher} {oxford university press},\ \bibinfo {year}
  {2013})\BibitemShut {NoStop}%
\bibitem [{\citenamefont {Read}, \citenamefont {Jagannathan},\ and\
  \citenamefont {Likhtman}(2008)}]{Read2008}%
  \BibitemOpen
  \bibfield  {author} {\bibinfo {author} {\bibfnamefont {D.~J.}\ \bibnamefont
  {Read}}, \bibinfo {author} {\bibfnamefont {K.}~\bibnamefont {Jagannathan}}, \
  and\ \bibinfo {author} {\bibfnamefont {A.~E.}\ \bibnamefont {Likhtman}},\
  }\bibfield  {title} {\enquote {\bibinfo {title} {Entangled polymers:
  Constraint release, mean paths, and tube bending energy},}\ }\href {\doibase
  10.1021/ma8009855} {\bibfield  {journal} {\bibinfo  {journal}
  {Macromolecules}\ }\textbf {\bibinfo {volume} {41}},\ \bibinfo {pages}
  {6843--6853} (\bibinfo {year} {2008})},\ \Eprint
  {http://arxiv.org/abs/http://dx.doi.org/10.1021/ma8009855}
  {http://dx.doi.org/10.1021/ma8009855} \BibitemShut {NoStop}%
\bibitem [{\citenamefont {Graham}\ \emph {et~al.}(2003)\citenamefont {Graham},
  \citenamefont {Likhtman}, \citenamefont {McLeish},\ and\ \citenamefont
  {Milner}}]{graham2003microscopic}%
  \BibitemOpen
  \bibfield  {author} {\bibinfo {author} {\bibfnamefont {R.~S.}\ \bibnamefont
  {Graham}}, \bibinfo {author} {\bibfnamefont {A.~E.}\ \bibnamefont
  {Likhtman}}, \bibinfo {author} {\bibfnamefont {T.~C.}\ \bibnamefont
  {McLeish}}, \ and\ \bibinfo {author} {\bibfnamefont {S.~T.}\ \bibnamefont
  {Milner}},\ }\bibfield  {title} {\enquote {\bibinfo {title} {Microscopic
  theory of linear, entangled polymer chains under rapid deformation including
  chain stretch and convective constraint release},}\ }\href@noop {} {\bibfield
   {journal} {\bibinfo  {journal} {Journal of Rheology (1978-present)}\
  }\textbf {\bibinfo {volume} {47}},\ \bibinfo {pages} {1171--1200} (\bibinfo
  {year} {2003})}\BibitemShut {NoStop}%
\bibitem [{\citenamefont {Likhtman}\ and\ \citenamefont
  {McLeish}(2002{\natexlab{a}})}]{likhtman2002quantitative}%
  \BibitemOpen
  \bibfield  {author} {\bibinfo {author} {\bibfnamefont {A.~E.}\ \bibnamefont
  {Likhtman}}\ and\ \bibinfo {author} {\bibfnamefont {T.~C.}\ \bibnamefont
  {McLeish}},\ }\bibfield  {title} {\enquote {\bibinfo {title} {Quantitative
  theory for linear dynamics of linear entangled polymers},}\ }\href@noop {}
  {\bibfield  {journal} {\bibinfo  {journal} {Macromolecules}\ }\textbf
  {\bibinfo {volume} {35}},\ \bibinfo {pages} {6332--6343} (\bibinfo {year}
  {2002}{\natexlab{a}})}\BibitemShut {NoStop}%
\bibitem [{\citenamefont {in~Supramolecular
  Polymer~Networks}(2013)}]{Seiffert2013}%
  \BibitemOpen
  \bibfield  {author} {\bibinfo {author} {\bibfnamefont {C.~D.}\ \bibnamefont
  {in~Supramolecular Polymer~Networks}},\ }\bibfield  {title} {\enquote
  {\bibinfo {title} {Hackelbusch, sebastian and rossow, torsten and van
  assenbergh, peter and seiffert, sebastian},}\ }\href {\doibase
  10.1021/acs.macromol.5b00733} {\bibfield  {journal} {\bibinfo  {journal}
  {Macromolecules}\ }\textbf {\bibinfo {volume} {46}},\ \bibinfo {pages}
  {6273--6286} (\bibinfo {year} {2013})},\ \Eprint
  {http://arxiv.org/abs/http://dx.doi.org/10.1021/ma4003648}
  {http://dx.doi.org/10.1021/ma4003648} \BibitemShut {NoStop}%
\bibitem [{\citenamefont {de~Gennes}(1982)}]{deGennesI1982}%
  \BibitemOpen
  \bibfield  {author} {\bibinfo {author} {\bibfnamefont {P.~G.}\ \bibnamefont
  {de~Gennes}},\ }\bibfield  {title} {\enquote {\bibinfo {title} {Kinetics of
  diffusion‐controlled processes in dense polymer systems. i. nonentangled
  regimes},}\ }\href {\doibase http://dx.doi.org/10.1063/1.443328} {\bibfield
  {journal} {\bibinfo  {journal} {The Journal of Chemical Physics}\ }\textbf
  {\bibinfo {volume} {76}},\ \bibinfo {pages} {3316--3321} (\bibinfo {year}
  {1982})}\BibitemShut {NoStop}%
\bibitem [{\citenamefont {Banerjee}, \citenamefont {Marchetti},\ and\
  \citenamefont {M{\"u}ller-Nedebock}(2011)}]{banerjee2011motor}%
  \BibitemOpen
  \bibfield  {author} {\bibinfo {author} {\bibfnamefont {S.}~\bibnamefont
  {Banerjee}}, \bibinfo {author} {\bibfnamefont {M.~C.}\ \bibnamefont
  {Marchetti}}, \ and\ \bibinfo {author} {\bibfnamefont {K.}~\bibnamefont
  {M{\"u}ller-Nedebock}},\ }\bibfield  {title} {\enquote {\bibinfo {title}
  {Motor-driven dynamics of cytoskeletal filaments in motility assays},}\
  }\href@noop {} {\bibfield  {journal} {\bibinfo  {journal} {Physical Review
  E}\ }\textbf {\bibinfo {volume} {84}},\ \bibinfo {pages} {011914} (\bibinfo
  {year} {2011})}\BibitemShut {NoStop}%
\bibitem [{\citenamefont {Milner}\ and\ \citenamefont
  {McLeish}(1998)}]{MilnerandMcLeish1998}%
  \BibitemOpen
  \bibfield  {author} {\bibinfo {author} {\bibfnamefont {S.~T.}\ \bibnamefont
  {Milner}}\ and\ \bibinfo {author} {\bibfnamefont {T.~C.~B.}\ \bibnamefont
  {McLeish}},\ }\bibfield  {title} {\enquote {\bibinfo {title} {Reptation and
  contour-length fluctuations in melts of linear polymers},}\ }\href {\doibase
  10.1103/PhysRevLett.81.725} {\bibfield  {journal} {\bibinfo  {journal} {Phys.
  Rev. Lett.}\ }\textbf {\bibinfo {volume} {81}},\ \bibinfo {pages} {725--728}
  (\bibinfo {year} {1998})}\BibitemShut {NoStop}%
\bibitem [{\citenamefont {Likhtman}\ and\ \citenamefont
  {McLeish}(2002{\natexlab{b}})}]{LikhtmanandMcLeish2002}%
  \BibitemOpen
  \bibfield  {author} {\bibinfo {author} {\bibfnamefont {A.~E.}\ \bibnamefont
  {Likhtman}}\ and\ \bibinfo {author} {\bibfnamefont {T.~C.~B.}\ \bibnamefont
  {McLeish}},\ }\bibfield  {title} {\enquote {\bibinfo {title} {Quantitative
  theory for linear dynamics of linear entangled polymers},}\ }\href {\doibase
  10.1021/ma0200219} {\bibfield  {journal} {\bibinfo  {journal}
  {Macromolecules}\ }\textbf {\bibinfo {volume} {35}},\ \bibinfo {pages}
  {6332--6343} (\bibinfo {year} {2002}{\natexlab{b}})}\BibitemShut {NoStop}%
\bibitem [{\citenamefont {Doi}\ and\ \citenamefont
  {Edwards}(1978)}]{DoiandEdwards1978}%
  \BibitemOpen
  \bibfield  {author} {\bibinfo {author} {\bibfnamefont {M.}~\bibnamefont
  {Doi}}\ and\ \bibinfo {author} {\bibfnamefont {S.~F.}\ \bibnamefont
  {Edwards}},\ }\bibfield  {title} {\enquote {\bibinfo {title} {Dynamics of
  concentrated polymer systems. part 2.-molecular motion under flow},}\ }\href
  {\doibase 10.1039/F29787401802} {\bibfield  {journal} {\bibinfo  {journal}
  {J. Chem. Soc.{,} Faraday Trans. 2}\ }\textbf {\bibinfo {volume} {74}},\
  \bibinfo {pages} {1802--1817} (\bibinfo {year} {1978})}\BibitemShut {NoStop}%
\end{thebibliography}%
\end{document}